\documentclass[letter]{aa} 
\usepackage{graphicx}
\usepackage{hyperref}
\usepackage{braket}
\usepackage{txfonts}
\usepackage{xcolor}
\definecolor{dark-red}{rgb}{0.9,0.0,0.0}
\definecolor{dark-blue}{rgb}{0.15,0.15,0.9}
\definecolor{dark-green}{rgb}{0.15,0.8,0.15}
\definecolor{medium-blue}{rgb}{0,0,0.9}
\hypersetup{
    colorlinks, linkcolor=red,
    citecolor={dark-blue} , urlcolor={medium-blue}
}
\begin{document} 

   \title{Empirical Contrast Model for High-Contrast Imaging}
   \subtitle{A VLT/SPHERE Case Study}
   \author{B. Courtney-Barrer\inst{1,2}, R. De Rosa\inst{1},R. Kokotanekova\inst{1,3},C. Romero\inst{1}, M. Jones\inst{1}, J. Milli\inst{1,4} \and Z. Wahhaj\inst{1}
          }
   \institute{European Southern Observatory, Alonso de C\'ordova 3107, Vitacura, Casilla 19001, Santiago, Chile\\
              \email{bcourtne@eso.org}
        \and
            Research School of Astronomy and Astrophysics, Australian National University, Canberra, ACT 2611, Australia
        \and
            Institute of Astronomy and National Astronomical Observatory, Bulgarian Academy of Sciences, 72 Tsarigradsko Shose Blvd., Sofia 1784, Bulgaria
        \and 
            Université Grenoble Alpes, CNRS, IPAG, 38000, Grenoble,France
             }
   \date{Received September 15, 1996; accepted March 16, 1997}

  \abstract
   {The ability to accurately predict the contrast achieved from high contrast imagers is important for efficient scheduling and quality control measures in modern observatories}
   {We aim to consistently predict and measure the raw contrast achieved by SPHERE/IRDIS on a frame by frame basis to improve the efficiency and scientific yield with SPHERE at the Very Large Telescope (VLT).}
   {Contrast curves were calculated for over 5 years of archival data using the most common SPHERE/IRDIS coronagraphic mode in the H2/H3 dual band filter, consisting of approximately 80,000 individual frames. These were merged and interpolated with atmospheric data to create a large data-base of contrast curves with associated features. An empirical power law model for contrast, motivated by physical considerations, was then trained and finally tested on an out-of-sample test data set.}
   {At an angular separation of 300 mas, the contrast model achieved a mean (out-of-sample) test error of 0.13 magnitudes with the residual 5-95\% percentiles between -0.23 and 0.64 magnitude respectively. The models test set root mean square error (RMSE) between 250-600 mas was between 0.31 - 0.40 magnitudes which is equivalent with other state-of-the-art contrast models presented in the literature. In general, the model performed best for targets between 5-9 G-band magnitude, with degraded performance for targets outside this range. This model is currently being incorporated into the Paranal SCUBA software for first level quality control and real time scheduling support.}
   {}
    \titlerunning{short title}
    \authorrunning{name(s) of author(s)}
   \maketitle
%
\section{Introduction}
    High contrast imagers have become central tools for the discovery and understanding of exoplanets and protoplanetary disk formation, demographics and dynamics around young stars. Instruments such as VLT's SPHERE \citep{Beuzit_2019_SPHERE, fusco_2015_saxo}, Gemini's GPI \citep{Macintosh_2014_GPIES_1st_light}, and Subaru's SCExAO \citep{Sahoo_2018_SCExAO} can achieve typical raw contrasts in the range of 10$^{-4}$ to 10$^{-6}$ between 0.1"-0.5" from the central star in near infra-red wavelengths. The foreseen instruments coming in the epoch of extremely large telescopes will push these boundaries even further (e.g. \citealt{Brandl_2021_METIS}). The outstanding capabilities for these current and future instruments attached to world class telescopes comes with high demand, making the optimisation of telescope time an important task. This optimization generally requires accurate models to predict observational performance indicators (e.g. contrast for high contrast imagers) both prior and/or early on during the observation in order to select observations that optimally exploit the atmospheric conditions. Traditionally, short term scheduling and quality control in queue observations for high contrast imagers such as SPHERE are done primarily based on atmospheric, sidereal or airmass constraints. While strong correlations exist between turbulence and AO performance, there are often outliers where the measured contrast are considerably worse than what would be expected from the observed atmospheric conditions. This is typically due to local effects within the telescope or instrument. Without the ability to properly predict and measure the contrast, such observations may be scheduled and pass basic quality control checks despite not meeting the users scientific requirements. Therefore, to optimise telescope time and the ultimate data quality provided to users, high contrast imagers may benefit greatly from precise models to predict scientifically meaningful metrics, such as contrast or SNR that can then be measured in quasi-real time to evaluate the quality of the data. The ultimate goal being to perform short-term scheduling and quality control based on predicted and measured metrics that hold scientific significance. This, combined with the significant efforts of improving quality control software and standards \citep{thomas_2020_scuba}, and forecasting models at Paranal \citep{Milli_2020_nowcast,milli_2019_nowcast,masciadri_2020_forecast_alg,osborn_2018_forecast_paranal}, will greatly advance real-time decision making and quality control measures. 
    The ability to accurately predict contrast for high contrast imagers on large telescopes is not a trivial task. While fundamental atmospheric limits are well characterized (\citealt{conan_1995, fusco_conan_2004_ao_stat, aime_2004_coronagraph_pinned_speckle, males_2021_speckle_lifetime}), typically non-trivial local effects can dominate achievable contrast, for example quasi-static speckles caused by opto-mechanical imperfections and thermal drifts (\citealt{bloemhof_2001, soummer_2007_speckle_noise, martinez_2013_quasi_static_speckle_sphere, vigan_zelda_2022}) or dome seeing (\citealt{tallis_2020_JATIS}) and low wind effects (\citealt{LWE1,LWE2}). Given the maturity of instruments such as SPHERE, data-driven analysis and empirical models are a practical way to understand these limitations and the uncertainty that random telescope/instrumental processes have on observations. Some good examples of this are  \citet{martinez_2013_quasi_static_speckle_sphere} work using SPHERE data to characterise speckle temporal stability in high Strehl regimes, \citet{LWE1} work characterising the low wind effect on SPHERE, and \citet{jones_2022_faint_targets} data-driven analysis of SPHERE performance for faint targets. In the case of predicting contrast - various models have been explored in literature to empirically predict the on-sky observed contrast given atmospheric and instrumental conditions \citep{baily_2016_perf_gpi,courtney_2019,xuan_2018_nirc2_contrast_pred}. In particular GPI's initial work using linear regression of AO telemetry and astronomical site monitoring data to predict contrast \citep{baily_2016_perf_gpi}, which was further advanced with neural networks that were able to predict the measured contrast using 6 input parameters that were available pre-observation with a contrast (magnitude) RMSE of 0.45 at 0.25" \citep{savransky_2018_minning_GPI}. Correlations between measured contrast and AO error terms has also been shown in other work (e.g. \cite{Poyneer_2006_speckle_behavior_in_high_contrast,Poyneer_2016_GPI_performance}), and in general have been shown to provide good predictive capacity of the contrast in atmospheric limited regimes \citep{SAXO1, SAXO2,Macintosh_2014_GPIES_1st_light}. For this work we present a simple empirical model to predict the raw contrast measured by SPHERE with the goal to assist on-site quality control measures and short-mid term scheduling decisions. This paper begins with a brief overview of the SPHERE instrument followed by motivating an empirical model to fit the contrast data to. Section 4 will outline the data preparation and pre-processing that was done before fitting the contrast model, along with the algorithms used for fitting. Section 5 will present the results along with some discussion. Section 6 will conclude our findings and future outlook.

\section{SPHERE/IRDIS}

The Spectro-Polarimetric High-contrast Exoplanet REsearcher (SPHERE) \citep{SPHERE_orig_paper} is an extreme adaptive optics (AO) instrument installed on the Unit Telescope 3 (Melipal) at the Paranal Observatory. Its primary science goal is imaging, low-resolution spectroscopic, and polarimetric characterization of extra-solar planetary systems at optical and near-infrared wavelengths. SPHERE consists of three science channels, the Integral Field Spectrograph (IFS) and the Infra-Red Dual-band Imager and Spectrograph (IRDIS), which both observe in the near-infrared, and the Zurich Imaging Polarimeter (ZIMPOL) for visible polarimetric observations. Each sub instrument has a series of coronagraphs and filters available in-addition to having an extreme AO system called SAXO \citep{SAXO1,SAXO2} placed in the common path of all sub instrument channels. SAXO operates up to a frequency of 1.38 kHz on bright targets with a 40x40 spatially filtered Shack-Hartmann (SH) wavefront sensor (WFS) measuring in the optical, and a 41x41 piezoelectric high-order deformable mirror for AO actuation. SAXO also uses a dedicated differential tip/tilt sensor \citep{Baudoz_2010_sphere_dif_tiptilt} in the near-infrared to correct for wavelength dependent tip tilt between the near-infrared and optical science channels. For this work we tested our model on the most common SPHERE/IRDIS mode which uses an apodised Lyot coronagraph with the H2/H3 dual band filters centered at wavelengths 1.593$\mu$m and 1.667$\mu$m respectively.

 \section{Contrast Model} \label{contrast_model}
We begin to motivate an empirical model for contrast with some statistical considerations of the measured intensity in the focal plane. It can easily be shown by Fourier optics that a phase aberration at some spatial frequency $k$ in a pupil plane of a telescope gets mapped to a so called speckle in the focal plane at an angular coordinate of $k \lambda$,  where $\lambda$ is the lights wavelength \citep{2004_roddier_textbook}. Such speckles are typically classified based on their temporal behaviour, which ultimately determines if (or how well) they can be suppressed by post processing reduction methods. This sets the fundamental contrast limits in ground based high contrast imagers \citep{males_2021_speckle_lifetime}. The detection of real signals (such as a planet) within the circumstellar environment of a star requires statistical knowledge on the probability of some intensity measurement in the focal plane. Various authors (e.g. \cite{canales_rician_dist_1999} and references therein) have shown under the  assumption of long exposures that a point wise intensity measurement (I) generally follows a modified Rician probability density function:
\begin{equation}
    P(I) = \frac{I}{2\sigma^2}\exp\left(-\frac{I+s^2}{\sigma^2}\right) I_0\left(\frac{2s\sqrt{I}}{\sigma^2}\right)
\end{equation}
where $I_0$ is the zero order modified Bessel function of the first kind. While $s^2$ and $2\sigma^2$ are related to the (long exposure) intensity of deterministic, and random speckle component of the wavefront respectively. \cite{soummer_2007_speckle_noise} developed this statistical framework to derive a general expression for the expected point wise variance in a coronagraphic image as: 
\begin{equation}
\sigma_I^2 = N(I_{s1}^2 + NI_{s2}^2 + 2I_{c}I_{s1} + 2NI_{c}I_{s2} + 2I_{s1}I_{s2} ) + \sigma_p
\label{eq:soummer_variance}
\end{equation}
where we have kept with the notation used in \citet{soummer_2007_speckle_noise}. Here 'I' generally denotes the intensity, $\sigma_p^2$ is the variance of the photon noise, and N is the ratio of fast-speckle and slow-speckle life times. $I_c$ is the intensity produced by the deterministic part of the wavefront, including static aberrations, while the $I_s$ terms correspond to the halo produced by random intensity variations, i.e. atmospheric ($I_{s1}$) and quasi-static contributions ($I_{s2}$).  In this generalized expression of the variance, several contributions can be identified by order of appearance: (1) the atmospheric halo; (2) the quasi-static halo; (3) the atmospheric pinning term, the speckle pinning of the static aberrations by the fast-evolving atmospheric speckles; (4) the speckle pinning of the static by quasi-static speckles; and finally (5) the speckle pinning of the atmospheric speckles by quasi-static speckles. Converting this to the expected contrast as a general function of radius (e.g. a typical contrast curve) requires calculating the sum of the pixel wise modified Rician density functions within a given annulus. No closed form analytic solution to this exists, although there are closed form approximations \citep{ricean_approx_2009}. Under a strong assumption of independence between pixels and, for a given spatial frequency, equal probability of the aberrations direction (i.e. angular position of a speckle at a fixed radius),  where $\theta$ is the angular position of speckle at given radius, we can estimate the expected intensity variance within a thin annulus at radius r by simply scaling by the number of pixels in the annulus, in which case we can make the proportional approximation of the 1$\sigma$ contrast:
\begin{equation}
\braket{C(r)} \propto \frac{\braket{\sigma_I(r)}}{I_*}
\label{eq:soummer_contrast}
\end{equation}
where $I_*$ is the stellar intensity in the science channel and $\braket{..}$ is the expected value. As mentioned, this is a strong assumption that does not generally hold. For example, experience shows that there is typically anisotropy in the aberrations at a given spatial frequency, especially from biased wind directions of dominant turbulent layers. Nevertheless this basic assumption is useful for deriving first order contrast estimates. Analytically predicting each term in \ref{eq:soummer_variance} prior to observation would require full knowledge of internal aberrations, wind velocity profiles and the ability to reconstruct modal distribution of incoming phase front in-order to predict AO residuals - which is a difficult task. Nevertheless, noting that any first order expansion of the coronagraph PSF term ($I_c$) and atmospheric speckle terms ($I_{s1}$) with regard to typical AO error budget terms would lead to various cross products of AO error budget terms in the pinned speckles; we could make the assumption that typically one of these terms will dominate the halo at a given radius and therefore propose to model the contrast as a product of AO cross terms, each with a power laws to give an appropriate weighting at a given radius. i.e. 
\begin{align}
C(r) &\approx x(r) \prod^{4}_{i=1} \Delta_i^{\alpha_i(r)}
\end{align}
where x(r) and $\alpha_i(r)$ are the fitted parameters for a given radius. From basic leave-one out analysis the $\Delta$ terms considered for the following model are a combination of typical (unitless) AO error budget like terms: 
\begin{align} \label{eq:zeta fitting parameters}
\Delta_1 &\equiv \Delta_{fit} = \frac{D}{r_0} \\
\Delta_2 &\equiv \Delta_{servo} = \frac{\tau}{\tau_0}  \\
\Delta_3 &\equiv \Delta_{SNR-WFS} = \frac{n_{p,wfs}}{\sqrt{n_{p,wfs}+N_D \left (n_B^2+\left( \frac{e_n}{G} \right) \right)}} \\
\Delta_4 &\equiv \Delta_{SNR-SCI} = \frac{1}{\sqrt{n_{p,sci}}}
\end{align}
where D is the telescope diameter, $r_0$ is the atmospheric coherence length (Fried parameter), $\tau$ and $\tau_0$ are the AO latency and atmospheric coherence time respectively, $n_p$ is the number of detected photoelectrons per defined subaperture (sum of all pixels); $N_D$ is the number of pixels in a subaperture, $n_B$ is the number of detected background photoelectrons per subaperture; $e_n$ is the read-noise in electrons per pixel, and G is the gain \citep{AO_field_guide}. The values $n_{p,.}$ are generally inferred through fitted zero points and extinction coefficients to convert stellar magnitude to flux (see section \ref{Flux_Calibrations}). The residuals of such a model would therefore be due to the variance in non-AO related terms in equation \ref{eq:soummer_variance}. We also note that by construction (also through cross validation on training data) in a shot noise limited regime the product of:
\begin{equation*}
\Delta_{red} \equiv  \Delta_{3}^{\alpha_{4}}\Delta_{4}^{\alpha_{4}} = n_{p,wfs}^{\alpha_3/2} / n_{p,sci}^{\alpha_4/2}
\end{equation*} 
can be seen as reddening parameter. In the case of equal flux $n_{p,wfs}=n_{p,sci}=n_p$ we get $\Delta_{red} = n_p^{(\alpha_{3}-\alpha_{4})/2} $ and therefore one would expect $\alpha_{4}>\alpha_{3}$ to maintain that brighter targets generally achieve better contrast. However in general $n_{p,wfs} \neq n_{p,sci}$ and chromatic effects of red stars may play an important role (especially with the performance of the differential tip-tilt controller). This general model has a considerable advantage that it can capture non-linearities in the contrast performance and furthermore can be fitted linearly by simply considering the contrast magnitude: $C_m=-5/2\log_{10}(C)$ such that: 
\begin{align}\label{eq: simplied mag contrast equation}
    C_m(r) = X(r) - \frac{5}{2}\sum^{4}_{i=1} \alpha_i(r) \log_{10}(\Delta_i)
\end{align}
where X(r) = $-5/2\log_{10}(x(r))$ is the fitted intercept. On the training dataset we also allowed for a linear calibration of the fitted intercept $X_f(r)$ with the model residuals given the partitioned instrumental state such that $X(r) = X_f(r) - \Delta_c(r|state)$. Where $\Delta_c(r|state)$ is the training model residual when filtering for a given observational state. The state filters considered were the sky transparency classification (e.g. thick, thin, clear, photometric) the AO gain/frequency setting, and the wavefront sensor spatial filter size (e.g. small, medium, large). This significantly improved the cross-validation performance of the model on the training data set, while maintaining a significant sample size for the general fitting of $\alpha$ parameters. These calibrated offsets were then used for the out-of-sample model test (without re-calibration).

\section{Data Preparation}
\begin{figure}[h]
\includegraphics[width=8cm]{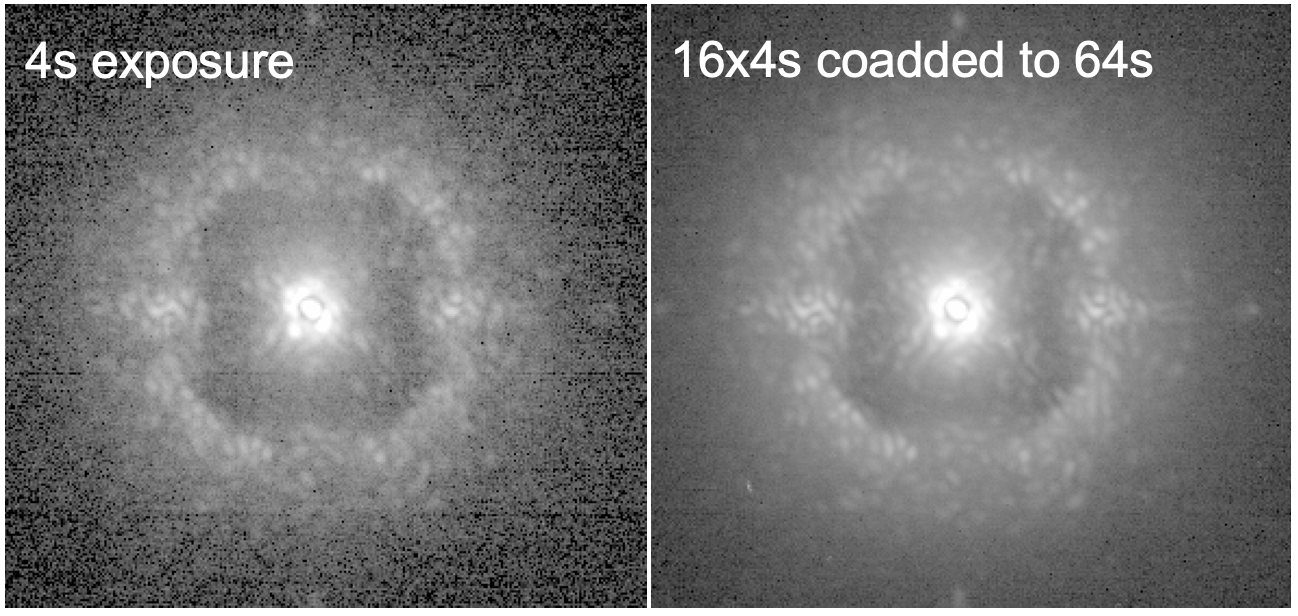}
\caption{An example of co-adding short exposure frames to 64s}   
\label{fig:coadding_frames}
\end{figure}

\begin{figure}[h]
\includegraphics[width=9cm]{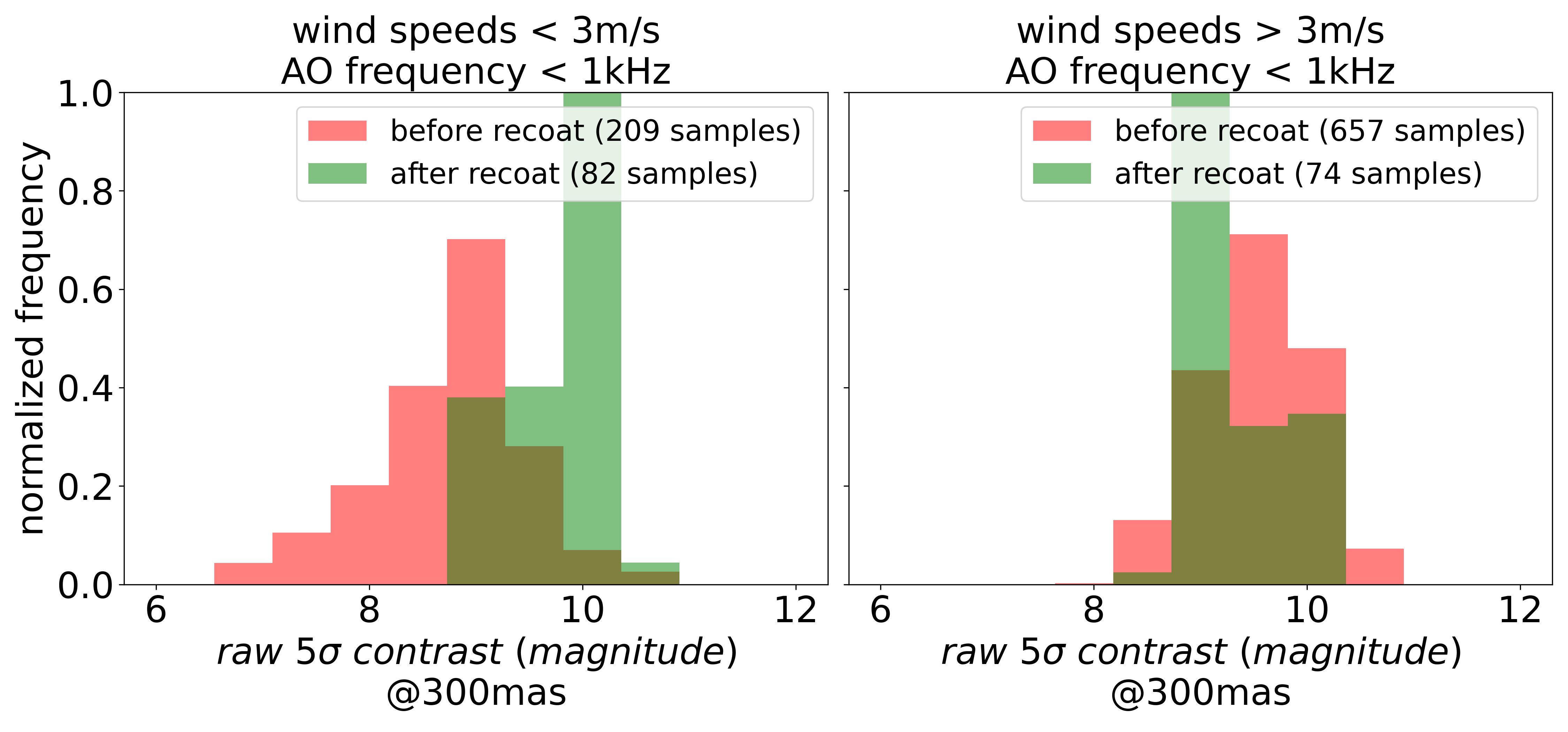}
\caption{Normalized histogram of raw contrast at 0.3" in [left] low wind conditions ($<3m/s$) and [right] nominal ($>3m/s$) wind conditions before (red) and after (green) M2 spider re-coat that was completed in November 2017. Fast AO modes (1200Hz, 1380Hz) were neglected in these histograms to avoid biases since the 1380Hz AO mode was an upgrade that was not available in earlier observations before the M2 spider re-coat.}   
\label{fig:m2_recoat}
\end{figure}

We developed a database of all public observations taken between 2015-2019 which were downloaded from the ESO SPHERE archive. This particular study focused on fitting the above described model for the most commonly observed SPHERE/IRDIS mode which uses an apodised Lyot coronapgraph with the H2 ($\lambda_c$ = 1.593$\mu$m) filter which corresponds to the left detector in the H2/H3 dual band mode. The general FITS headers used to filter this data are displayed in table \ref{tab:keywords}. After filtering and outlier rejection (described below) train (75\%) and test (25\%) data sets were split to have non-overlapping observation nights - meaning for any given sample in the train set, there did not exist a sample in the test set that was observed on the same night (and vis versa). This corresponded to 149 and 47 unique stars in the train and test sets respectively with only 4 stars shared between the two sets, totalling to nearly 80k raw coronagraphic frames to analyse. A 75/25\% split provided a sufficient parameter space density to perform 10-fold cross validation on the training set, while allowing sufficient samples to avoid biases in the out-of-sample test.
\begin{table*}
\caption{Header keywords used to filter the data for the coronagraphic observations. \label{tab:keywords}}
\centering
\begin{tabular}{lccc}
\hline\hline
\vspace{-0.3cm} \\
Keyword          & Value   \\
\hline \vspace{-0.3cm} \\
DPR TYPE         &  OBJECT / OBJECT,FLUX                \\
INS COMB ICOR   &  N\_ALC\_YJH\_S    \\
DPR TECH         &  IMAGE    \\
INS1 FILT NAME    &  B\_H                    \\ 
INS COMB IFLT   &  DB\_H23   \\ 
INS4 FILT3 NAME   &  OPEN                    \\
\hline 
\end{tabular}
\end{table*}
1$\sigma$ noise levels were estimated as a function of radius in coronagraphic data cubes (DPR TYPE = OBJECT) after some basic reduction (e.g. background subtraction, flat fielding, bad pixel masking and high-pass filtering). The standard deviation was calculated in an annulus with 4 pixel ($\sim\lambda/D$) from 82-1800mas, where radii that had pixels in non-linear regime of IRDIS detector (ADU > 20k) were masked. Additionally, each coronagraphic (OBJECT) frame was cross correlated with median coronagraphic image across the filtered data set to provide an additional criteria for outlier removal. From visual inspection of individual frames, anything with a  cross correlation below  0.5 seemed to correspond to frames that had obvious issues such as bright companions in the field or where AO loops temporarily opened during an exposure. Therefore frames that had a cross correlation with the median image less than 0.5 were dropped from the analysis. In short exposures there were noticeable pinned speckles that were initially difficult predict from atmospheric conditions. This was significantly improved by co-adding coronagraph frames that had exposure times $\leq$ 64s to roughly 64s. An example of this is shown in figure \ref{fig:coadding_frames}.
A 2D Gaussian function was then fitted to the corresponding non-coronagraphic flux frames (DPR TYPE = OBJECT,FLUX) of the observation to estimate the peak flux. The contrast curve was then calculated by dividing the 1$\sigma$ noise levels at a given radii by the estimated the peak flux when adjusting for the different integration times and neutral density filters. To correct for changing transparency between the non-coronagraphic (flux) and coronagraphic (object) frames, the measured contrast was multiplied by the ratio of aggregated wavefront sensor (SPARTA telemetry) flux data between the two periods. While no specific sky classifications (e.g. thin or thick clouds) were excluded from the data, we neglected data where there was significant ($>50$\%) variability in the wavefront sensor flux during an exposure, since this caused significant variability in the measured contrast that was unpredictable from a pre-observation perspective.
\\\\
The initial contrast curve database consisted of 27135 co-added contrast frames. Each contrast curve was associated with various features including the FITS headers from the initial files, and the full available range of atmospheric parameters available from ESO MASS-DIMM and meteorological archives which were interpolated to the mean coronagraphic frame timestamp. This included important parameters such as atmospheric seeing and coherence time. Atmospheric data prior to the last MASS-DIMM upgrade (April 2016) were neglected due to instrumental biases between the old and new systems. Data was then further filtered to exclude observations that were outside of standard operational conditions and/or where feature outliers were detected. Basic filters included that:
\begin{itemize}
\item All SPARTA AO loops were closed
\item SPARTA differential tip/tilt loop is closed
\item Telescope was guiding on a guidestar
\item Atmospheric seeing and coherence time measurements were in the range 0-5" and 0-30ms respectively.
\item Raw coronagraphic image cross-correlation with median image $>$ 0.5
\item No low wind effected data (typically V$<$3m/s in data before Nov-2017)
\item wavefront sensor flux variability did not exceed more than 50\% between frames.
\end{itemize}
Data taken prior to the M2 spider re-coating done in November 2017 shows significantly different contrast statistics for wind speeds below 3m/s due to the low wind effect, which was largely fixed by the re-coating intervention \citep{LWE1,LWE2}. Figure \ref{fig:m2_recoat} shows the measured raw contrast at 0.3" before and after the M2 spider re-coat for low (left plot) and nominal (right plot) wind speeds. It is clear that there was a large statistical improvement in the measured contrast from the intervention in the low wind case, while a statistically insignificant difference when observing in wind speeds $>$3m/s after the re-coating. Note that data taken with fast AO modes (1200Hz, 1380Hz) were neglected in these histograms to avoid biases since the 1380Hz AO mode was an upgrade that was not used in earlier data, particularly before the M2 spider re-coat.  This prompted us to only neglect data taken before November 2017 with wind speed $<$3m/s. After this filtering process 8494 co-added frames remained for training and testing the model.


\subsection{Flux Model}\label{Flux_Calibrations}
Calibrated instrumental zero points and extinction coefficients were required to estimate the photocurrent (ADU/s) recieved in both the wavefront sensor and science detector for a given SIMBAD magnitude and airmass prior to observation. Data was first filtered to consider SPHERE flux sequences and WFS data taken during periods classified as photometric using either the LP780 or OPEN spectral filter without restrictions on the WFS spatial filter size. From this filtered data the following model was fitted:
\begin{align}
    M=\beta_0 \left (2.5 \log_{10}{\left (\frac{F}{T G}\right )} \right )+\beta_1 X+\beta_2
\end{align}
Where M is the Simbad magnitude of the target star, F is the flux (ADU / s) measured in either the WFS (G band) which is summed over all sub apertures, or flux template from the science detector (H band), T is a scalar to represent the relative transmission of the spectral filter, G the detector gain,  X is the targets airmass,  and $\beta_0, \beta_1, \beta_2$ are the fitted parameters corresponding to the telescope/instrument transfer function, extinction coefficient and zero point respectively. Fitted parameters from data taken in photometric conditions are outlined in table \ref{tab:flux_fit} and are consistent with previously measured extinction coefficients at Paranal \citep{Patat_2011_extinction_paranal}.
\\
\begin{table}
 \caption{\label{tab:flux_fit} Fitted parameters for the WFS (G band) and Science flux frame (H band) magnitude to flux model in photometric conditions.}
  \centering
    \begin{tabular}{ |c||c|c|  } 
     \hline
     Parameter & WFS (G-Band) & H2 FILTER \\
     \hline
     $\beta_0$   & -1.057    & -0.899\\
     $\beta_1$&   0.127  & 0.147\\
     $\beta_2$& 25.260 & 17.705 \\
     \hline
    \end{tabular}
\end{table}
\\\\
To account for sky transparency in the contrast model sky category offsets were calibrated on the train data set via the contrast residuals of the respective sky category partitioned data as described in section \ref{contrast_model}.
\begin{figure}[h]
\includegraphics[width=18cm]{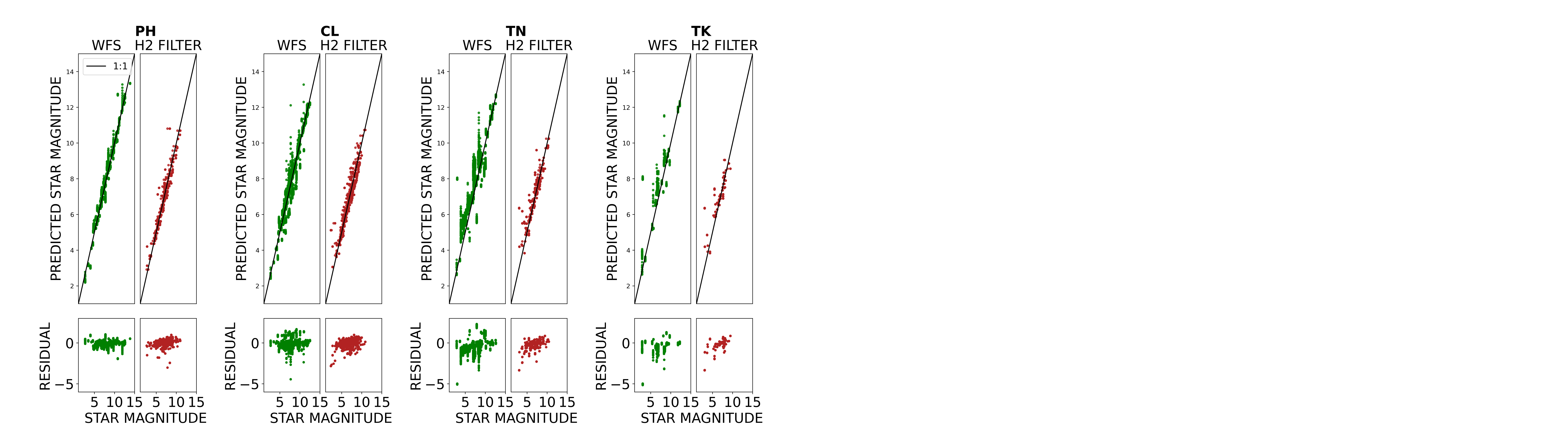}
\caption{G and H band photometric models applied to different sky transparency categories defined as Photometric (PH), Clear (CL), Thin (TN), Thick (TH)}   
\label{fig:photometric_models}
\end{figure}
\begin{figure}[h]
\includegraphics[width=10cm]{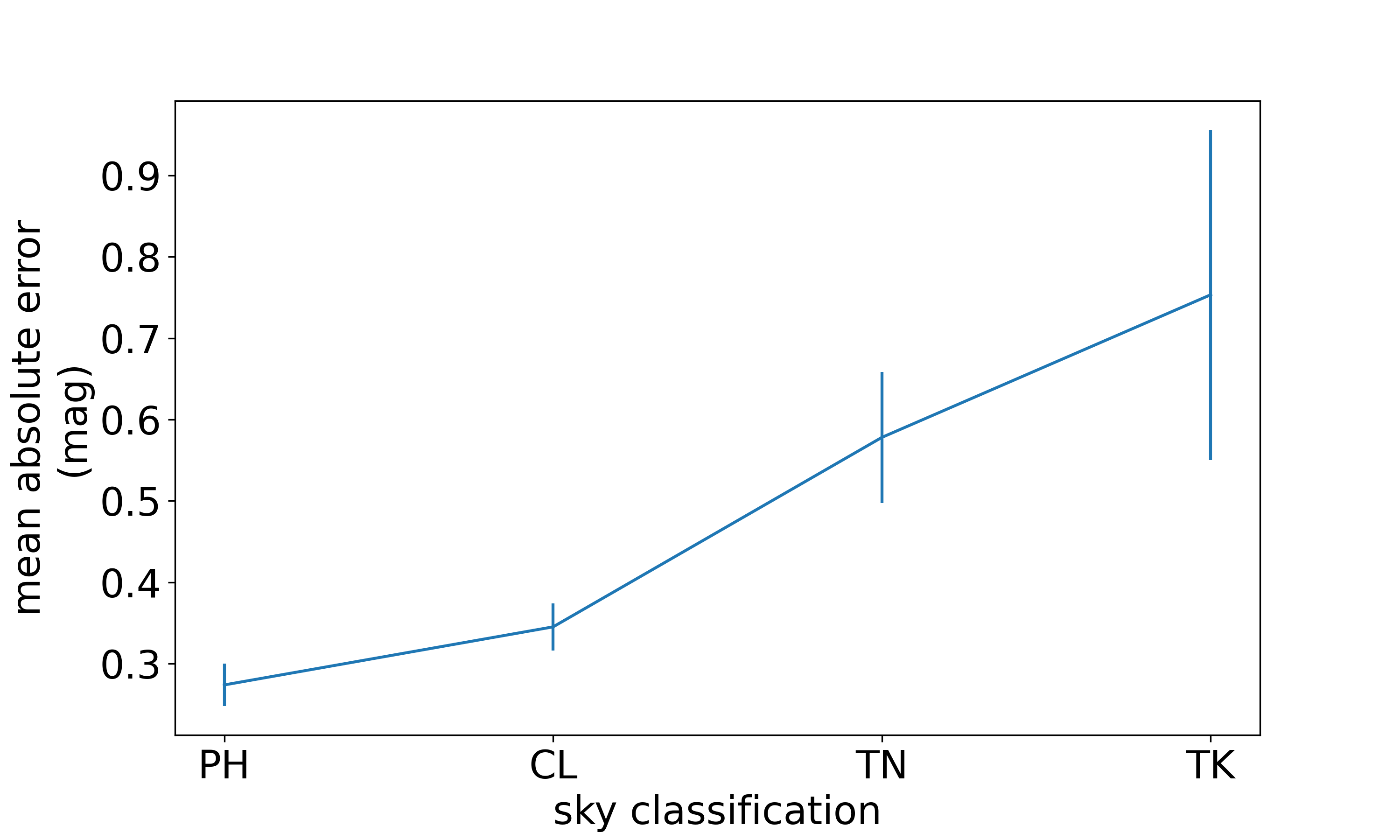}
\caption{Mean absolute error in the WFS flux model vs weather officer sky classification}  
\label{fig:model_residuals_v_sky_class}
\end{figure}
Additionally figure \ref{fig:photometric_models} displays the results when the data was partitioned into sky-transparency categories as classified by the weather officer and the above fitted photometric model applied to each respective sky category. Figure \ref{fig:model_residuals_v_sky_class} shows that the mean absolute error between the measured and predicted WFS flux given the target magnitude scales monotonically with the weathers officer classification of the sky transparency. These results suggest that the models could be used for automatic sky transparency classification. This would be advantageous over the standard method of the weather officer going outside every 30 minutes to visually classify the whole sky, since the WFS measurements would be in real-time directly within the SPHERE field of view.

\subsection{Model Fitting}
The empirical contrast model presented in section \ref{contrast_model}  was fitted with the python scikit-learn package \cite{scikit-learn} using Bayesian regression. This model was tuned and ultimately fitted on the training data set (75\% of the data) using 10-fold cross validation. The best fit parameters are reported as the mean of the 10-fold fit on the training set for the given radius, and respective uncertainties are $\pm$ two standard deviations. After tuning via cross validation on the training data set, the model evaluated on the test data set (25\% of the data). The results are presented in the following sections.


\section{Results and Discussion}
\begin{figure}[h]
\includegraphics[width=16cm]{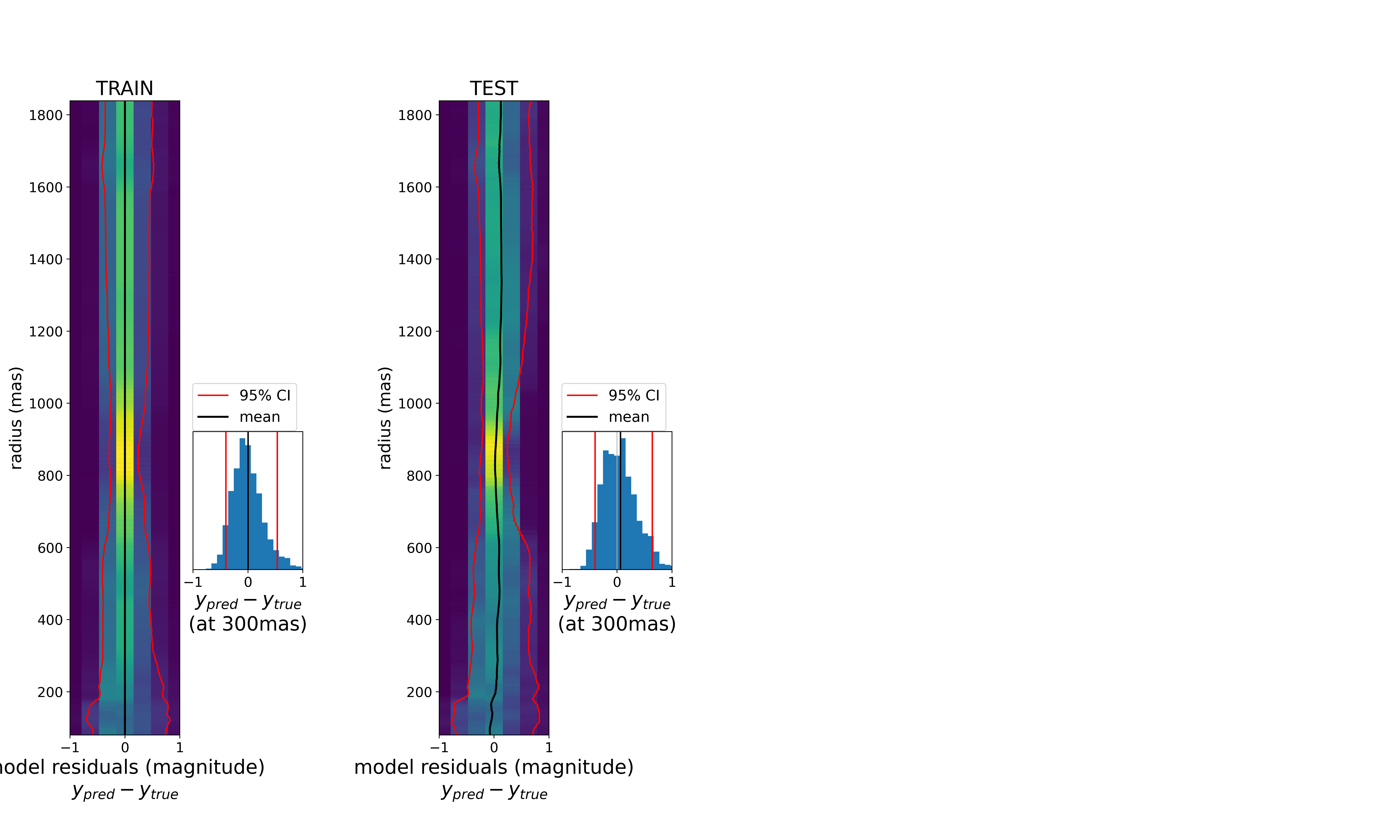}
\caption{Train and test contrast curve residual heatmaps (2D histograms) with sample 1D histograms shown at a 300mas radius}   
\label{fig:model residuals}
\end{figure}

\begin{figure}[h]
\includegraphics[width=10cm]{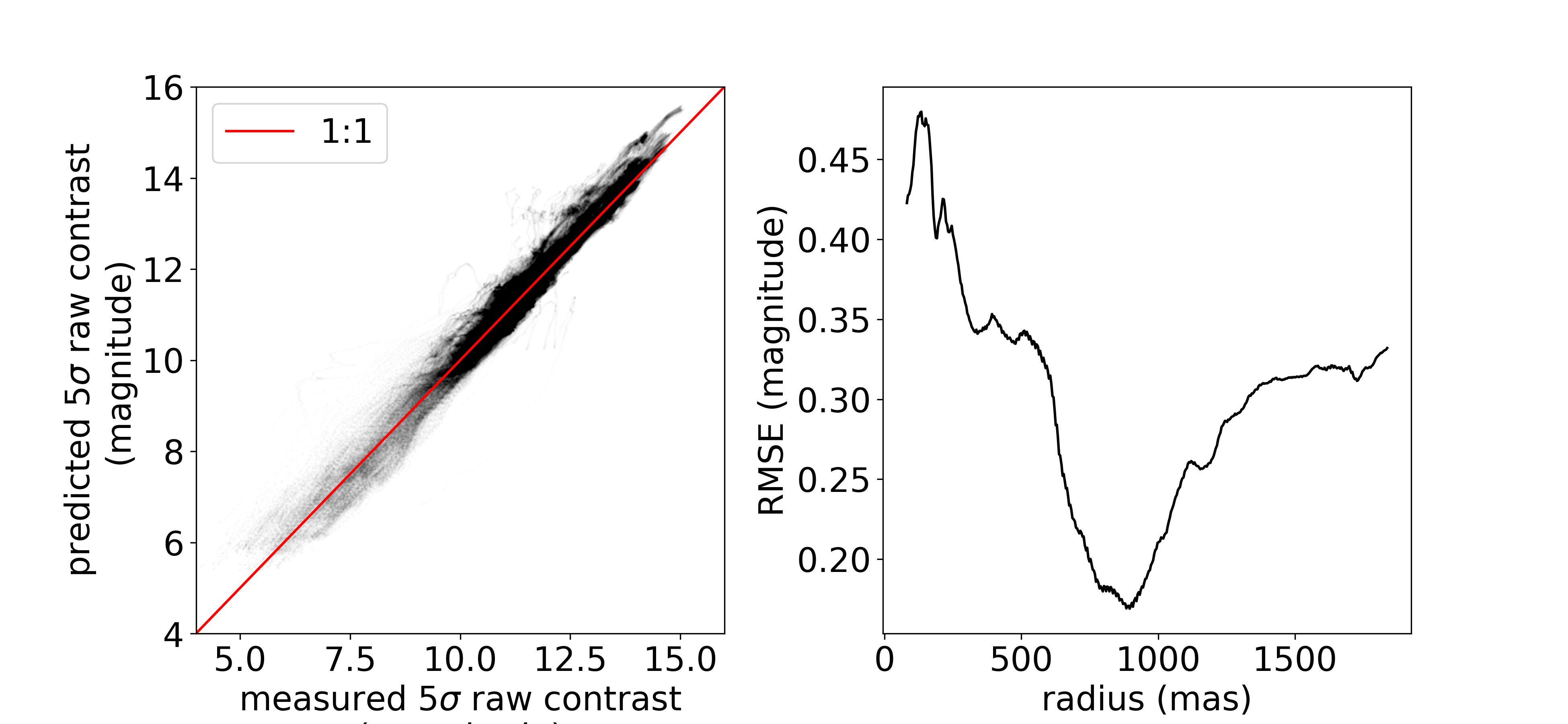}
\caption{Test set results. [Left] measured vs predicted raw 5$\sigma$ contrasts plotted in magnitudes. [Right] RMSE for the raw 5$\sigma$ contrast magnitude vs radius}   
\label{fig:test rmse}
\end{figure}

Figure \ref{fig:model residuals} shows the residuals of the contrast model as a function of separation from the central star for the train and test data sets, which shows good generalization to the out of sample test. It can also be seen that mean model residuals were well centered near zero, with 0.13 magnitude mean error, and residual 5-95\% percentiles between -0.23, 0.64 magnitude respectively on the test set at 300mas. Figure \ref{fig:test rmse} shows the general measured vs predicted raw $5\sigma$ contrast on the out-of-sample test set across all radii, along with the respective model RMSE at a given radial separations from the central star. Between 0.25"-1.00" the model achieved a RMSE between 0.17-0.40 magnitude on the out-of-sample test set.

\begin{figure*}
\includegraphics[width=30cm]{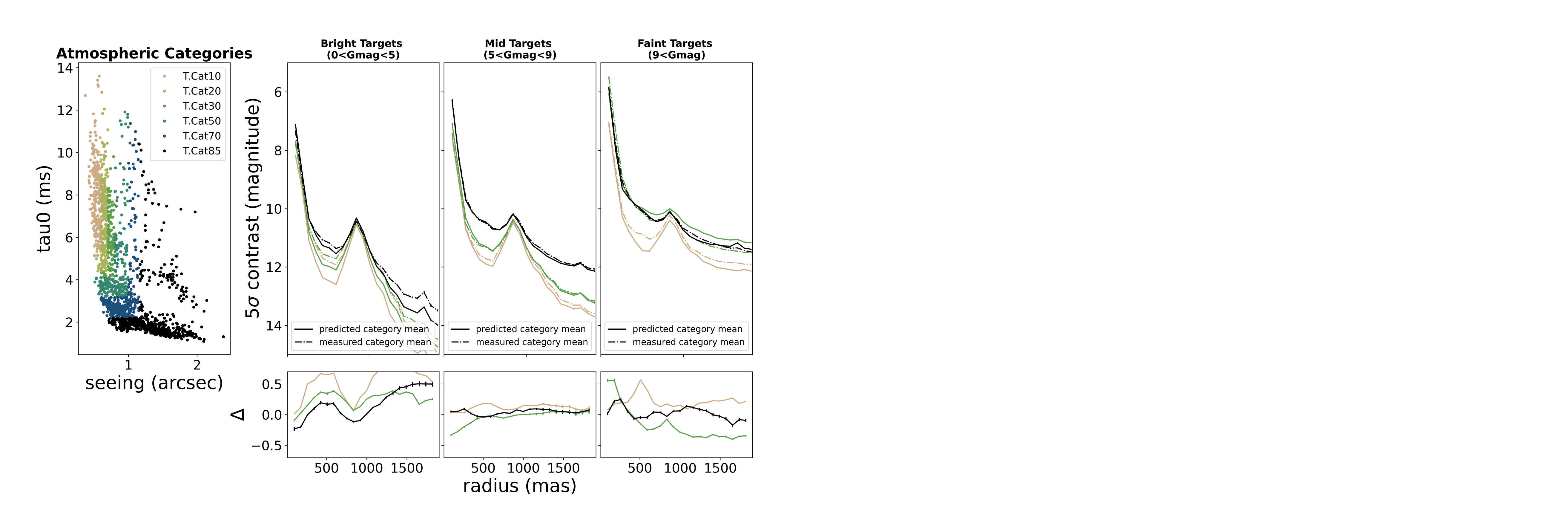}
\caption{Test data predicted vs measured contrast binned by star magnitude (faint, mid, bright) and also the atmospheric categories currently used for ranking and grading SPHERE observations in Paranal}   
\label{fig:test_atm_mag_bin_contrast}
\end{figure*}

\begin{figure}[h]
\includegraphics[width=8cm]{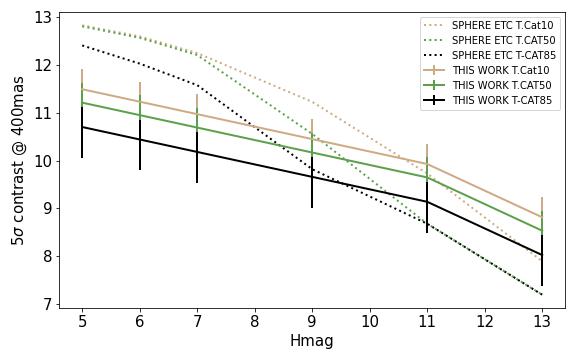}
\caption{Comparison between the 400mas contrast predictions made by the model of this work vs the current (as of publication of this paper) ETC on the ESO website. The ETC options were set to match the observed mode of the model developed here. i.e. DBH23 filter with cornagraph. Both the ETC and model were set with no neutral densities, exptime$\sim$64s, DIT$\leq$64s, without differential imaging and considered a spectral type around GV2}   
\label{fig:etc_comparison}
\end{figure}

\begin{figure}[h]
\includegraphics[width=8cm]{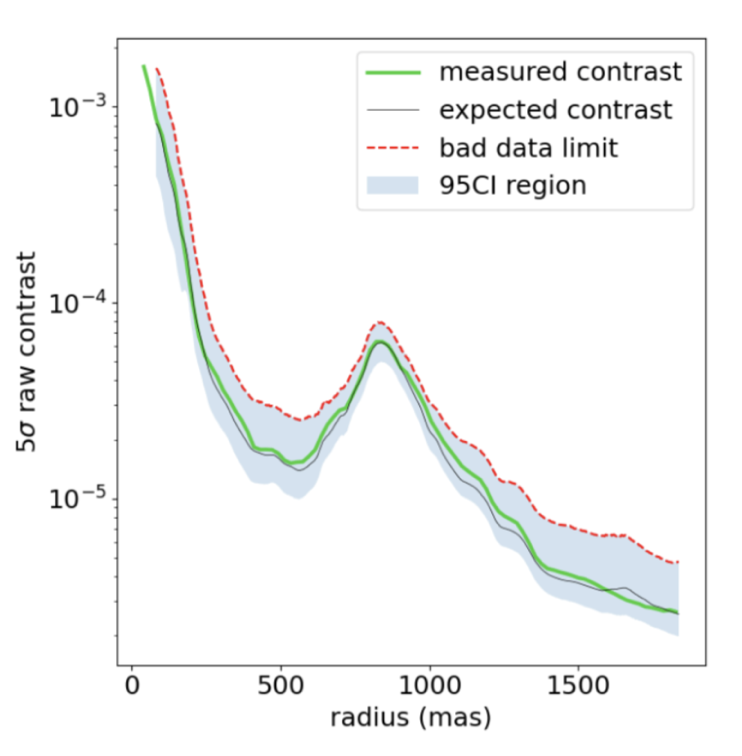}
\caption{Example of how the contrast model could be used in operations for providing quick checks to ensure the measured raw contrast ($\sim$60s frame) is within the expected 95\% range of model residuals given the target and current atmospheric conditions.}   
\label{fig:model_in_ops}
\end{figure}

We also analyze the model performance over an aggregated grid of atmospheric and star brightness categories in figure \ref{fig:test_atm_mag_bin_contrast}. The atmospheric categories considered are those currently used (ESO period p106) as user constraints for grading observations. The categories are defined from cuts in the atmospheric seeing, coherence time joint probability distribution, with T.Cat10 corresponding to the best (top 10\%) atmospheric conditions, and T.cat85 to the worst. We also simply consider 3 star magnitude categories of bright (Gmag < 5), mid (5<Gmag<9) and faint (Gmag>9) targets. There is excellent agreement (<0.15mag residual at 0.5") in the mid category range across all atmospheric conditions, however relatively worse performance for the bright and faint categories, particularly when in better atmospheric conditions. This would indicate that the underlying assumption that pinned atmospheric residuals dominates contrast is most valid in the mid category, while worst performance is seen for bright and faint target where, for example, static/quasi-static pinning may become dominant.  
\begin{figure}[h]
\includegraphics[width=8cm]{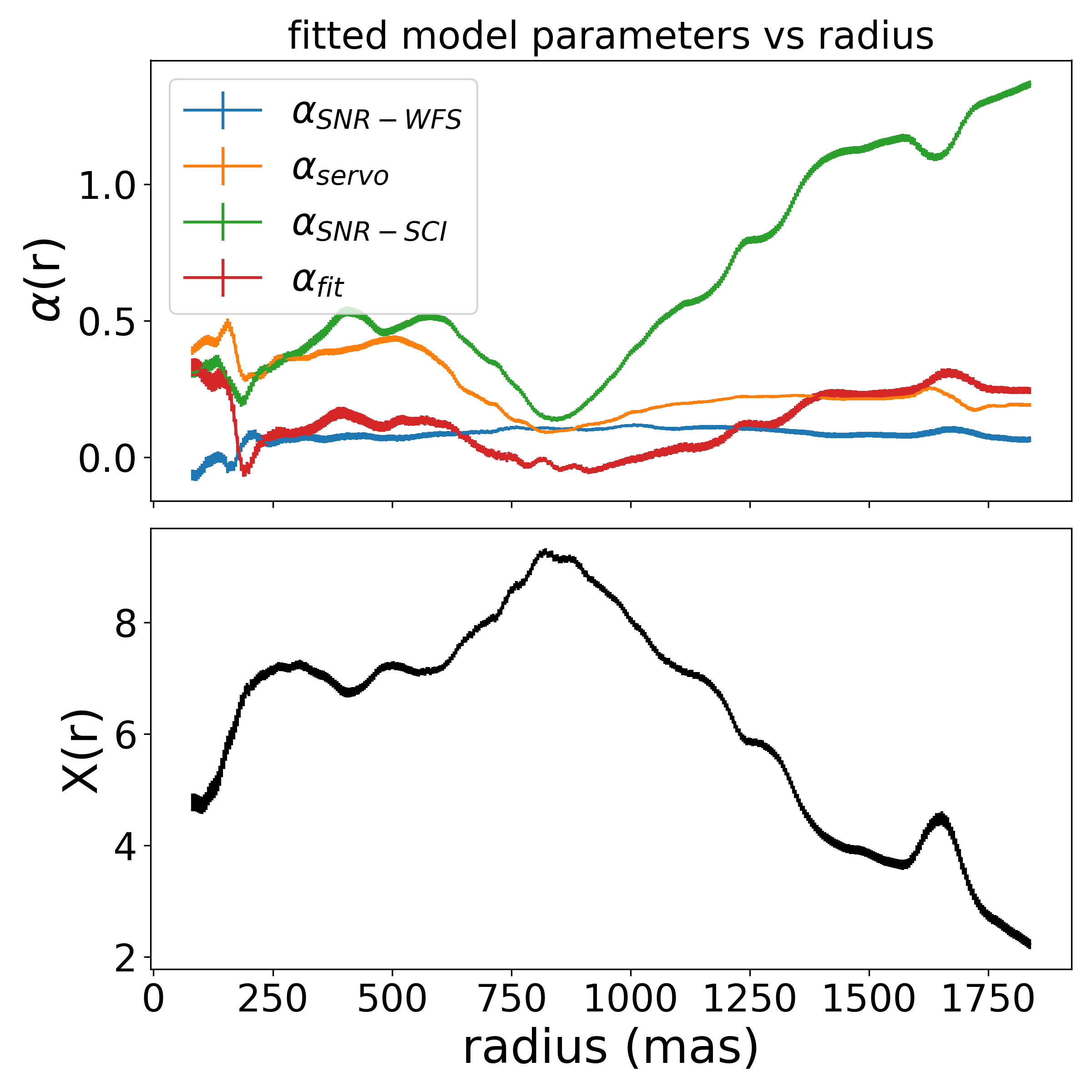}
\caption{[Top] The mean fitted alpha power indicies $\pm2\sigma$ as a function of separation.  [Bottom] The fitted intercept $X(r)$ (in log space) as a function of separation.}   
\label{fig:fitted_coes}
\end{figure}
The fitted parameters are shown in figure \ref{fig:fitted_coes}. We find fitted power indicies deviate considerably from the 5/3 power laws typically encountered in AO error budgets for phase residuals arising from limited spatial or temporal  bandwidths. Between 250-500mas these typically range between 10-30\% the 5/3 value for fitting errors and servo lag related terms. The fitted parameter typically approach zero (minimum sensitivity) near a radius of 800mas. This corresponds to the radius of the  which is determined by the inter-actuator spacing of the deformable mirror. Interestingly the fitting term has two zero crossing points, going negative (albeit very near zero) between 750-1000mas which is around the scattering halo. This implies that contrast within this radii slightly degrades with lower seeing. Outside of this region we get the expected behaviour that contrast improves with lower seeing. However, it is clear that SPHERE contrast is much more sensitive to the atmospheric coherence time rather than coherence length (seeing). For example, at 300mas doubling the atmospheric coherence time (keeping all other variables equal) leads to an expected $\sim30$\% reduction (improvement) in contrast, while doubling the Fried parameter (halving the seeing) only leads to a $\sim6$\% reduction in contrast. Similar results were also found for the Gemini Planet Imager \citep{baily_2016_perf_gpi}. As expected $\alpha_{SNR-WFS} < \alpha_{SNR-SCI}$ for all radii which, as discussed in section \ref{contrast_model} implies that contrast generally improves with brightness assuming equal partitioning of flux between WFS and science channels in a shot noise limited regime. Analysing the reddening parameter we see that contrast generally degrades as targets become more red. For example, around 300mas considering the train set median Gmag=7, contrast degrades by roughly 20\% per magnitude difference between WFS and Science channels. 
\\\\
To compare results to other contrast models found in literature, we achieved a test contrast RMSE between 0.31 - 0.40 magnitudes between 250 - 600mas, or equivalently a log10 contrast RMSE of 0.13-0.16 respectively. This is comparable to the results of  \citet{savransky_2018_minning_GPI} which, when using a feedforward neural network with pre-observation data as input, achieved a test log10 contrast RMSE of 0.18 at 250mas. Such comparable results are encouraging given the relative simplicity and physical interpretability of the model presented in this work compared to more complex neural networks. Comparing the predictions of this model at 400mas to the current SPHERE exposure time calculator (ETC) offered by ESO (as of May 2023) when considering raw image predictions (EXPTIME~64s) without differential imagining (neglecting field rotation) we see in figure \ref{fig:etc_comparison} that the ETC appears to provide very optimistic predictions for bright targets, and pessimistic for faint targets relative to the predictions of this work's model across all turbulence categories. This model also predicts that the contrast is less sensitive to changes in H-band flux compared to the the ETC which seems to also be reflected in the data at hand. The inclusion of this data to the ETC in short exposure time limits could be used to ultimately improve predictive accuracy of an ETC for SPHERE users. The contrast model presented here is currently being incorporated into Paranal's SCUBA software \citep{thomas_2020_scuba} to be used as first level quality control and help improve the real-time decision process for SPHERE at Paranal. 
Figure \ref{fig:model_in_ops} shows a real example of how this model could be used in operations for providing quick checks to ensure the measured raw contrast ($\sim$60s frame) is within the expected 95\% test residual range of the model given the target and current atmospheric conditions. Statistics on the frame-by-frame contrast could then be used to grade the OB based on potential user constraints. Abnormal aberrations caused from instrumental effects that impact the contrast can also easily be detected and evaluated by users and/or operators in the context of expected performance. Based on the out-of-sample test results; at 0.3" the operator should be able to predict the contrast to less than 0.5 magnitude at a 2 sigma level. 
\\\\

\section{Conclusions}
A simple product of power laws model was trained and tested on SPHERE/IRDIS coronagraphic data to predict contrast as a function of radius in the most commonly used H-band filter. When testing on out-of-sample test data, the model had a mean error of 0.13 magnitudes with residual 5-95\% percentiles between -0.23 and 0.64 magnitude respectively at 300mas. The models test set RMSE between 250-600mas was between 0.31 - 0.40 magnitudes which was on-par with other state-of-the-art contrast models presented in literature. This model is currently being incorporated into the Paranal SCUBA software for first level quality control and real time scheduling support. Future work will consider fitting these model to other SPHERE instrumental modes. 



\begin{acknowledgements}
This work has made use of the the SPHERE Data Centre, jointly operated by OSUG/IPAG (Grenoble), PYTHEAS/LAM/CESAM (Marseille), OCA/Lagrange (Nice), Observatoire de Paris/LESIA (Paris), and Observatoire de Lyon. 

\end{acknowledgements}

%
%

\bibliographystyle{aa}

\bibliography{sphere_bib}

\end{document}